\documentclass[sigconf]{acmart}

\usepackage{amsmath,amssymb,amsfonts}
\usepackage{algorithmic}
\usepackage{textcomp}
\usepackage{xcolor}
\usepackage{graphicx}
\usepackage{url}
\usepackage{dsfont}
\usepackage{soul}
\usepackage[T1]{fontenc}
\usepackage{listings}
\usepackage{float}

\lstdefinelanguage{PythonCustom}{
  keywords={
    def,return,if,else,elif,for,while,break,continue,
    import,from,as,class,try,except,finally,with,lambda,
    True,False,None
  },
  keywordstyle=\color{green!60!black}\bfseries,
  comment=[l]\#,
  commentstyle=\color{gray}\ttfamily,
  stringstyle=\color{red},
  sensitive=true,
  morekeywords=[2]{}, 
  keywordstyle=[2]\color{blue}\bfseries
}

\def\eqref#1{Equation~\ref{#1}}

\begin{document}

\copyrightyear{2026}
\acmYear{2026}
\setcopyright{usgov}
\acmConference[SCA/HPCAsiaWS 2026]{SCA/HPCAsia 2026 Workshops: Supercomputing Asia and International Conference on High Performance Computing in Asia Pacific Region Workshops}{January 26--29, 2026}{Osaka, Japan}
\acmBooktitle{SCA/HPCAsia 2026 Workshops: Supercomputing Asia and International Conference on High Performance Computing in Asia Pacific Region Workshops (SCA/HPCAsiaWS 2026), January 26--29, 2026, Osaka, Japan}
\acmPrice{}
\acmDOI{10.1145/3784828.3785240}
\acmISBN{979-8-4007-2328-5/2026/01}

\title{Q-IRIS: The Evolution of the IRIS Task-Based Runtime to Enable Classical-Quantum Workflows}

\newcommand{\equalcontrib}{NRM, MAHM, and EW are equal contributors and designated as co-first authors.}

\author{Narasinga Rao Miniskar}
\authornote{\equalcontrib}
\affiliation{
  \institution{Oak Ridge National Laboratory}
  \city{Oak Ridge}
  \state{TN}
  \country{USA}
}
\email{miniskarnr@ornl.gov}

\author{Mohammand A.H. Monil}
\authornotemark[1] 
\affiliation{
  \institution{Oak Ridge National Laboratory}
  \city{Oak Ridge}
  \state{TN}
  \country{USA}
}
\email{monilm@ornl.gov}

\author{Elaine Wong}
\authornotemark[1]
\affiliation{
  \institution{Oak Ridge National Laboratory}
  \city{Oak Ridge}
  \state{TN}
  \country{USA}
}
\email{wongey@ornl.gov}

\author{Vicente Leyton-Ortega}
\affiliation{
  \institution{Oak Ridge National Laboratory}
  \city{Oak Ridge}
  \state{TN}
  \country{USA}
}
\email{leytonorteva@ornl.gov}

\author{Jeffrey S. Vetter}
\affiliation{
  \institution{Oak Ridge National Laboratory}
  \city{Oak Ridge}
  \state{TN}
  \country{USA}
}
\email{vetter@ornl.gov}

\author{Seth R. Johnson}
\affiliation{
  \institution{Oak Ridge National Laboratory}
  \city{Oak Ridge}
  \state{TN}
  \country{USA}
}
\email{johnsonsr@ornl.gov}

\author{Travis S.~Humble}
\affiliation{
  \institution{Oak Ridge National Laboratory}
  \city{Oak Ridge}
  \state{TN}
  \country{USA}
}
\email{humblets@ornl.gov}

\renewcommand{\shortauthors}{Narasinga Rao Miniskar, Mohammad A.H. Monil, Elaine Wong, et al.}


\begin{abstract}

Extreme heterogeneity in emerging HPC systems are starting to include quantum accelerators, motivating runtimes that can coordinate between classical and quantum workloads. We present a proof-of-concept hybrid execution framework integrating the IRIS asynchronous task-based runtime with the XACC quantum programming framework via the Quantum Intermediate Representation Execution Engine (QIR-EE). IRIS orchestrates multiple programs written in the quantum intermediate representation (QIR) across heterogeneous backends (including multiple quantum simulators), enabling concurrent execution of classical and quantum tasks. Although not a performance study, we report measurable outcomes through the successful asynchronous scheduling and execution of multiple quantum workloads. To illustrate practical runtime implications, we decompose a four-qubit circuit into smaller subcircuits through a process known as quantum circuit cutting, reducing per-task quantum simulation load and demonstrating how task granularity can improve simulator throughput and reduce queueing behavior—effects directly relevant to early quantum hardware environments. We conclude by outlining key challenges for scaling hybrid runtimes, including coordinated scheduling, classical–quantum interaction management, and support for diverse backend resources in heterogeneous systems.
\end{abstract}

\maketitle

\section{Introduction}

Harnessing the potential of quantum computers to accelerate software applications requires hybrid programming paradigms to express and execute operations across conventional and quantum platforms. While the quantum software community has been developing full-stack solutions for dedicated quantum computing hardware, the essential concern of integrating such paradigms into existing software architectures, frameworks, and tools~\cite{pennylane2022,qiskit2024,pyquil2016} has been largely unaddressed. This concern recognizes both the integral nature of conventional methods for existing applications and the expectation that quantum and classical computing models must operate concurrently. Moreover, integration of quantum computing with asynchronous, many-task frameworks will depend heavily on the granularity of the quantum computing tasks, and methods that can expose such programming constructs to the application developer will be highly valued for expressiveness and sophistication.

As an example, consider that the implementation of quantum algorithms for a quantum computer must invariably require instantiation and initialization through a conventional programming interface, i.e., a classical control system that has the responsibility of storing, launching, and managing execution of the quantum program~\cite{quam2024,qick2022,qubic2023}. In a given application context, the initialization of the quantum algorithms are nearly always dependent on additional classical computation in which subsequent processing and analysis of the returned results demands further processing. Even the simplest example of a quantum circuit to prepare a two-qubit entangled Bell state will require input that specifies the state to be prepared and output that is processed to validate the presence of entanglement. Generally, the integration of quantum computing with asynchronous HPC workflows required for more complex hybrid methods is even more sophisticated.

Alongside the above hybrid computational challenges, there is also a pressing need to support heterogeneous computation in the quantum computing context. Presently, there is wide diversity of quantum computing hardware under consideration in which many different technologies are being used to test and evaluate the performance and behavior of applications~\cite{arute2019superconductingsupremacy,bourassa2021faulttolerantphotonic,DARPAQBI,henriet2020neutralatoms,trappedions,loss1998quantumdots,siddiqi2021highcoherencesuperconducting,zhong2020photonicadvantage}. Notwithstanding the ability of a full-stack environment to support an individual hardware technology, the portability of programs across different hardware platforms often suffers due to a lack of standardization in the programming languages and compiler as well as a strong reliance on platform-specific customizations to reach acceptable outcomes. One such approach has been the introduction of a hardware-agnostic Quantum Intermediate Representation (QIR)~\cite{qirspec}, serving as an abstraction layer between quantum programs and hardware. Moreover, the QIR Alliance~\cite{QIRAlliance} was formed to guide the development of QIR by providing tools to enable portable hybrid execution that contributes to the efforts for benchmarking, reliability, and reproducibility in quantum computer science. The QIR Execution Engine~\cite{qireerepo,wong2025qiree} is one such tool, and provides interfaces for processing and executing QIR code. However, executions of multiple programs are serial in nature and any attempt to simultaneously run a large set of quantum subprograms (which may or may not utilize classical compute tools) would ultimately be hindered by design.

Accordingly, we look to known, robust, classical frameworks for inspiration on how to manage parallel executions effectively and to determine whether or not quantum components can be similarly managed. To this end, we present some initial results for a portable hybrid programming paradigm based on extension of the IRIS~\cite{kim2024iris,miniskar2024iris} heterogeneous runtime system to accommodate parallel, asynchronous quantum computation. Our approach leverages the task-based programming model from IRIS alongside the recently developed hardware-agnostic QIR~\cite{qirspec} and its associated execution environment QIR-EE~\cite{qireerepo,wong2025qiree}. 
Encouragingly, the modular nature of both IRIS and QIR-EE facilitates the possibility of designing algorithms from the outset to capitalize simultaneously on both the classical and quantum backends.

\section{Background}
\label{sec:background}

\subsection{IRIS: Intelligent Runtime System}

IRIS~\cite{kim2024iris,miniskar2024iris} is a heterogeneous runtime system with the support of diverse heterogeneous devices such as CPUs, GPUs (Nvidia and AMD), FPGAs (Xilinx and Intel) and DSPs (Snapdragon Hexagon). It provides simplified task programming model APIs to create and submit tasks with the kernels written in heterogeneous programming models such as CUDA for Nvidia GPUs, HIP for AMD GPUs, OpenCL for any CPU and GPUs, Xilinx high level synthesis (HLS) C++ for FPGAs, DSP C++ for Hexagon DSPs. IRIS has intelligent capabilities: automatic data orchestration~\cite{miniskar2023iris} to move data across heterogeneous devices and host memory; configurable task scheduling policies~\cite{johnston2024iris} to decide task mappings dynamically to run on heterogeneous devices; and, dynamic compilation that allows lazy evaluation of kernels. IRIS has been used as a runtime backend for high-level programming models such as OpenACC, OpenACC, SYCL, Python and Julia.
IRIS is modular and extendable for new compute units such as QPUs (Quantum Processing Units) with domain specific programming models such as QIR.  In this paper, we add quantum compute capabilities to IRIS with a goal to achieve high portability and productivity to run the quantum algorithms on CPUs, GPUs and QPUs. 

\subsection{QIR-EE: Quantum Intermediate Representation Execution Engine}

The Quantum Intermediate Representation Execution Engine (QIR-EE) is a piece of ORNL developed software~\cite{qireerepo,wong2025qiree} written in C++ that enables the parsing of classical-quantum programs written in the Quantum Intermediate Representation (QIR)~\cite{qirspec} for the purpose of accelerating them to be executed on hardware and simulators. QIR was initially developed with the intention of promoting interoperability between different frameworks and languages and to easily connect emerging technologies for researchers to experiment with distinct and differentiated hardware capabilities.

As such, the ability to execute QIR programs on a diverse set of backends became an immediate concern that was solved by QIR-EE. After building the library, the executable can be invoked, receiving inputs such as the *.ll file (containing the program expressed in QIR), choice of accelerator, and number of shots. This enables the exceution of the QIR file on the backend dictated by the accelerator, and provides an easy access point for QIR-EE to be used as a runtime for the quantum processing unit (QPU) in the IRIS workflow.

The QIR-EE implementation relies heavily on the LLVM runtime to handle the processing of classical instructions while presenting interfaces to facilitate bindings with backend functions and capabilities. Some simulators and hardware access are currently provided through the XACC~\cite{xaccrepo} framework. However, what has been lacking in the QIR-EE stack is the ability to process multiple programs on multiple accelerators asynchronously. The integration with the IRIS framework is a first proof-of-concept step towards orchestrating this type of execution for QIR programs.
\section{Q-IRIS Design}
\label{sec:qiris}

This section introduces Q-IRIS, which integrates quantum computation capabilities in the asynchronous task-based execution paradigm of heterogeneous runtime IRIS. The purpose of Q-IRIS is to introduce the notion of QPU capability into the IRIS ecosystem as a quantum analogue to CPUs and GPUs, where classical computation can be carried out on CPUs and GPUs, and quantum computation can be scheduled on quantum hardware or simulators. A key challenge with including quantum computation capability in a heterogeneous task-based execution paradigm is the creation of necessary software infrastructure where a runtime scheduler can asynchronously execute a task involving quantum devices, while allowing the classical computation task to be executed asynchronously and simultaneously. Q-IRIS strives to provide a solution for this challenge by introducing the notion of a ``quantum task'' in the IRIS runtime ecosystem. We organize the section as follows: the Q-IRIS software stack is discussed in~\ref{sec:qiris-stack}, where the IRIS runtime software stack is enriched with quantum execution capabilities via QIR-EE. Then, we describe two variants of the IRIS programming model in Sections~\ref{sec:python-wrapper} and~\ref{sec:task-graph} and compare their scaling potential with a toy example in Section~\ref{sec:ghz}. We provide some outlook and a roadmap for future work in Section~\ref{sec:outlook}.

\begin{figure*}[!ht]
    \centering
    \vspace{-0.1in}    \includegraphics[width=0.8\textwidth,scale=1.0,keepaspectratio]{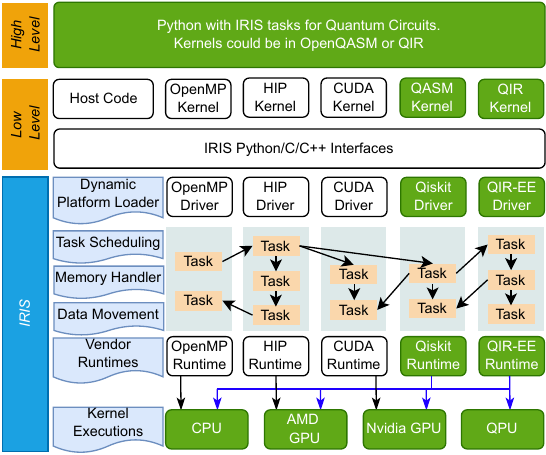}
    \caption{Q-IRIS software stack shows IRIS runtime with quantum capabilities through Qiskit runtime driven OpenQASM/QASM kernels and QIR-EE driven QIR kernels. QIR-EE can run the QIR kernels either using QSim or XACC.}
    \label{fig:irisq}
    \vspace{-0.1in}
\end{figure*}

\subsection{Q-IRIS Software Stack}
\label{sec:qiris-stack}
The Q-IRIS software stack is presented in Fig.~\ref{fig:irisq}, which includes the execution of quantum tasks through QASM and QIR kernels. At a high level, the IRIS programming model allows one to express the task code (along with the dependencies between the tasks) in Python, C, or C++ APIs. IRIS tasks specify only the kernel names and their required memory objects. IRIS uses the abstracted heterogeneous decentralized memory model (DMEM) and orchestrates the automatic movement of data between devices based on the mapping of tasks to the device and data availability. At run-time, the dynamic platform loader probes the availability of a device in a heterogeneous system and dynamically loads the shared objects of the devices drivers to facilitate the kernel launch, device memory allocation, data transfers, context, and stream management. Q-IRIS includes the Qiskit and QIR-EE drivers' shared objects so that the quantum task kernels can be invoked by those driver APIs. The current schedulers of the IRIS runtime support asynchronous execution of tasks by respecting the task dependencies, which are indicated by arrows in Fig.~\ref{fig:irisq}. When a task is scheduled for a particular device, IRIS interacts with the vendor-provided library to launch computation and communication on the device or accelerator. Examples of runtimes are CUDA, HIP, etc.

\subsubsection{QIR-EE as a Runtime}
\label{sec:runtime}
To realize the full potential of task-based execution in IRIS, Q-IRIS considers QIR-EE as a runtime (like CUDA, HIP or OpenCL) to include seamless execution of quantum tasks along with classical tasks. In other words, QIR-EE, as an execution manager, can be viewed as a runtime environment when connected to the right backend to enable execution of the QIR program. This kind of connection is presented as an `app' in QIR-EE's implementation, where instructions are mapped according to the backend requirements.

In Q-IRIS, the usage of XACC ~\cite{xaccrepo} as an app plugged into QIR-EE opens up the XACC toolbox, with access to a variety of backends that are available there. As of this writing, this includes the Aer~\cite{qiskit2024} and QPP~\cite{QuantumPP} (Quantum++) simulators, as well as access to vendor hardware backends from ionq, quantinuum, and ibm. This is not a restriction, as other apps such as Google's qsim and Xanadu's lightning~\cite{XanaduLightning} have been implemented independent of XACC within QIR-EE's framework. Since the IRIS runtime has the capability to concurrently manage multiple device-specific runtimes, considering QIR-EE as a runtime enables Q-IRIS to seamlessly and simultaneously execute classical and quantum tasks.

In the following sections, two programming approaches of task-based execution of quantum computation in IRIS runtime and their advantages and disadvantages are discussed.

\begin{figure}[htp]
\centering
\begin{lstlisting}[language=Python]
import iris
import os
def qiree_task(iris_params, iris_dev):

    params = iris.iris2py(iris_params)
    dev = iris.iris2py_dev(iris_dev)
    
    # QIR-EE Inputs
    execpath = params[0]
    filepath = params[1]
    accelerator = params[2]
    numshots = params[3]

    # QIR-EE Execution
    results = os.system(execpath+" "+filepath+" -a "+accelerator+" -s "+numshots+" > output.txt")
    
    return iris.iris_ok

# Begin parallel tasks here
iris.init()

# Parameters should be a list of strings in the following order:
parameters = ["executable", "LL file", "name of accelerator to use", "number of shots"]

# Running in parallel
n = 16

# Create n tasks (no communication with each other)
tasks = [iris.task() for i in range(n)]

# Feed parameters to each task and submit
# sync tells you how long to wait for completion of the command
for i in range(n):
    tasks[i].pyhost(qiree_task, parameters)
    tasks[i].submit(iris.iris_default, sync=0)

iris.synchronize()
iris.finalize()
\end{lstlisting}
\caption{Python wrapper for QIR-EE task to run 16 parallel executions of a QIR program on quantum hardware.}
\label{fig:python-wrapper}
\end{figure}

\begin{figure}[t]
\centering 
\begin{lstlisting}[language=Python]
import iris
import qiskit
from qiskit import QuantumCircuit
from qiskit_aer import AerSimulator

def go_quantum(iris_params, iris_dev):

    params = iris.iris2py(iris_params)
    dev = iris.iris2py_dev(iris_dev)
    
    circ_qasm = params[0]
    
    meas = QuantumCircuit(2, 2)
    meas.measure(range(2), range(2))
    qc = meas.compose(circ_qasm, range(2), front=True)
    backend = AerSimulator()
    job_sim = backend.run(qc, shots=1024)
    result_sim = job_sim.result()
    counts = result_sim.get_counts(qc)
        
    return iris.iris_ok
\end{lstlisting}
\caption{Q-IRIS via Qiskit Kernel
}
\label{fig:task-qiskit}
\end{figure}

\subsection{Python Wrapper}
\label{sec:python-wrapper}

One straightforward way to integrate QIR-EE with IRIS is to construct a python function that executes the QIR-EE command given input parameters as a list of string consisting of the paths to the executable and QIR file, together with accelerator choice (e.g., ibm's AER simulator) and number of shots (e.g., 1024). IRIS tasks kernels can then be constructed as a list with the \textsc{iris.task()} command and relayed to host memory. Figure~\ref{fig:python-wrapper} gives the pseudocode for this wrapper and a sample of what a single task execution would look like.

For this first ``proof-of-concept'' parallel execution on quantum processing units (QPUs), there is no dependency between tasks. In principle, the tasks themselves can include different programs to be executed on different quantum backends that is facilitated by QIR-EE. These backends may include quantum simulators or hardware, whatever QIR-EE has access to. The tasks are submitted to IRIS to enable the execution of the QIR-EE commands. Beyond QIR-EE, a separate driver could be loaded to showcase another example of QPU parallelism, provided that a corresponding runtime is available to access backends. This can be realized via IBM's Qiskit runtime: as long as the circuit and transpilation libraries are loaded, the construction of a task wrapper utilizing Qiskit commands can replace the ones for QIR-EE as in Figure~\ref{fig:python-wrapper}. As a result, using IRIS for parallel accelerations to multiple QPUs can potentially permit execution of hybrid programs in a quantum-heterogeneous way. While this approach is convenient for rapid prototyping, it does not harness the true capabilities of asynchronous task-based runtime IRIS.

\begin{figure}[htp]
\centering 
\begin{lstlisting}[language=Python]
#!/usr/bin/env python3

import iris
import numpy as np
import sys
iris.init()
SIZE = 8 if len(sys.argv) == 1 else int(sys.argv[1])
# Task creation and definition of QIR kernel bell.ll
task0 = iris.task()
task0.kernel("bell.ll", 1, [], [SIZE], [], [] , [] )
# Graph of task creation where multiple dependent tasks can exist
graph = iris.graph([task0])
graph.submit()
graph.wait()
iris.finalize()
\end{lstlisting}
\caption{Quantum task creation in IRIS. Here, one IRIS task is created for the QIR-EE back-end of IRIS; however, other tasks can be added by specifying the dependencies, enabling quantum and classical execution simultaneously.}
\label{fig:task-qiree}
\end{figure}

\subsection{IRIS Task Abstraction for Quantum Kernels}
\label{sec:task-graph}
To enable simultaneous execution of quantum and classical tasks, Q-IRIS uses QIR-EE as a back end, similar to the CUDA or HIP back end of the IRIS runtime. Having QIR-EE as the back-end, Q-IRIS enables the preservation of architecture-agnostic semantics of the IRIS task code. Figure~\ref{fig:task-qiree} shows the creation of architecture-agnostic tasks for the execution of the QIR kernel. Line 9 provides the corresponding QIR kernel where the name of the kernel is $bell.ll$. Note that this kernel execution only resolves at run-time. One can provide a CUDA or HIP kernel implementation with the same name and will be able to execute on CUDA or HIP GPUs. At runtime, IRIS runtime schedules $task0$ on quantum hardware or simulator using the QIR-EE runtime, where $bell.ll$ is passed to QIR-EE to be executed through XACC. The difference between this approach and the wrapper approach is that more tasks along with their dependencies can be created. Such a set of tasks with their dependencies can be scheduled through the IRIS task scheduler, potentially enabling simultaneous and asynchronous execution of quantum and classical tasks based on the task dependencies. Although this approach allows us to realize the full task-based execution capabilities of the IRIS runtime, it required the enhancement of IRIS runtime by creating a quantum device backend (similar to the GPU backend of IRIS) that invokes QIR-EE. Having such a QIR-EE backend in IRIS runtime, a user can create a regular IRIS host program to execute through QIR-EE by providing the necessary QIR files.

\section{GHZ Quantum State Preparation}
\label{sec:ghz}

\begin{figure}[h]
    \centering
    \includegraphics[width=\linewidth]{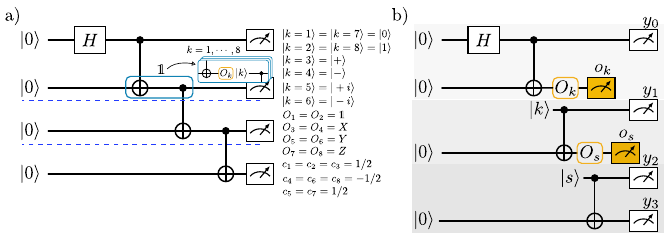}
    \caption{GHZ preparation using a QDP with two quantum wire cuts. Panel (a) illustrates the quantum circuit for the four-qubit GHZ state preparation, with wire cuts depicted as dashed blue lines. This panel details the decomposition of the quantum wire into multiple quantum instances, presenting the values for the coefficients and corresponding elements for the QDP introduced in Eq. \ref{eq:qdp} \cite{harada2024doubly}. Panel (b) provides a clearer visualization of each quantum instance depicted with a different shade of gray, each composed of three two-qubit quantum circuits. }
    \label{fig:ghz}
\end{figure}

In this section, we demonstrate Q-IRIS capabilities by presenting a parallel execution of quantum simulations. For this illustrative example, we focus on the preparation of a four-qubit Greenberger-Horne-Zeilinger (GHZ) state~\cite{Greenberger1989,computing2000quantum}, a resource in quantum computing for illustrating quantum entanglement.  

The GHZ state for four qubits is defined as $(|0000\rangle + |1111\rangle) / \sqrt{2}$, representing a quantum superposition where the four qubits are either in the $|0000\rangle$ or in the $|1111\rangle$ state with equal probability. This state exhibits a maximal entanglement, meaning they can be perfectly correlated -- measuring any single qubit immediately determines the state of the others. Preparing a maximally entangled state involves a series of entangling gates that can challenge both quantum hardware and simulation software due to quantum-specific constraints. Unlike classical computation, where complexity is typically measured in terms of time and space, quantum complexity also depends on factors such as circuit depth, qubit connectivity, and decoherence effects (imperfections) \cite{computing2000quantum}.  

With the goal to show a parallel execution, the GHZ preparation quantum circuit is decomposed into a set a of two-qubit quantum circuits through a quasi-probability decomposition (QPD)~\cite{howard2017application,pashayan2015estimating,seddon2021quantifying}. This approach not only simplifies the quantum simulation load but also leverage the parallel computation capabilities on HPC resources.

Following the decomposition, we implement two quantum wire cuts represented by the blue dashed lines in Fig.~\ref{fig:ghz}a, which splits the quantum circuit into smaller segments. In this  approach~\cite{harada2024doubly}, the quantum wire, represented as an identity channel $\mathds{1}$ that preserves the quantum state, i.e., $\mathds{1} | \varphi \rangle \langle \varphi | = | \varphi \rangle \langle \varphi | $, is expressed by the combination
\begin{equation} \label{eq:qdp}
    \mathds{1} (| \varphi \rangle \langle \varphi |) = \sum_{k=1}^8 c_k {\rm Tr}[O_k (| \varphi \rangle \langle \varphi |)] |k \rangle \langle k | \, ,
\end{equation} 
where $O_k \in \{ \mathds{1}, X, Y, Z\}$ is acting on an ancillary qubit and \[|k \rangle \in \{ | 0\rangle , |1\rangle , |+\rangle, |+i\rangle , |-i\rangle \}\] is a new initial quantum state on the target qubit where $\mathds{1}$ is acting. In the decomposition (\ref{eq:qdp}), ${\rm Tr}[O_k (| \varphi \rangle \langle \varphi|)] |k \rangle \langle k |$ represents a channel that measures the expectation value of the observable $O_k$ on the quantum state $|\varphi \rangle$ entering the identity channel $\mathds{1}$ and prepares a new quantum state $|k \rangle$. In Fig.~\ref{fig:ghz}a we  show the quantum circuit for the GHZ quantum state preparation, and more details for the QPD. For two quantum wire cuts the decomposition generates a batch of $8 \times 8$ quantum instances, each consisting of three quantum circuits (distinguished by different shades of gray in Fig.~\ref{fig:ghz}b). These quantum instances are distributed across 64 cores. Each core independently executes one quantum instance using either the qiskit runtime or QIR-EE. 

\begin{figure}[h]
    \centering
    \includegraphics[width=0.9\linewidth]{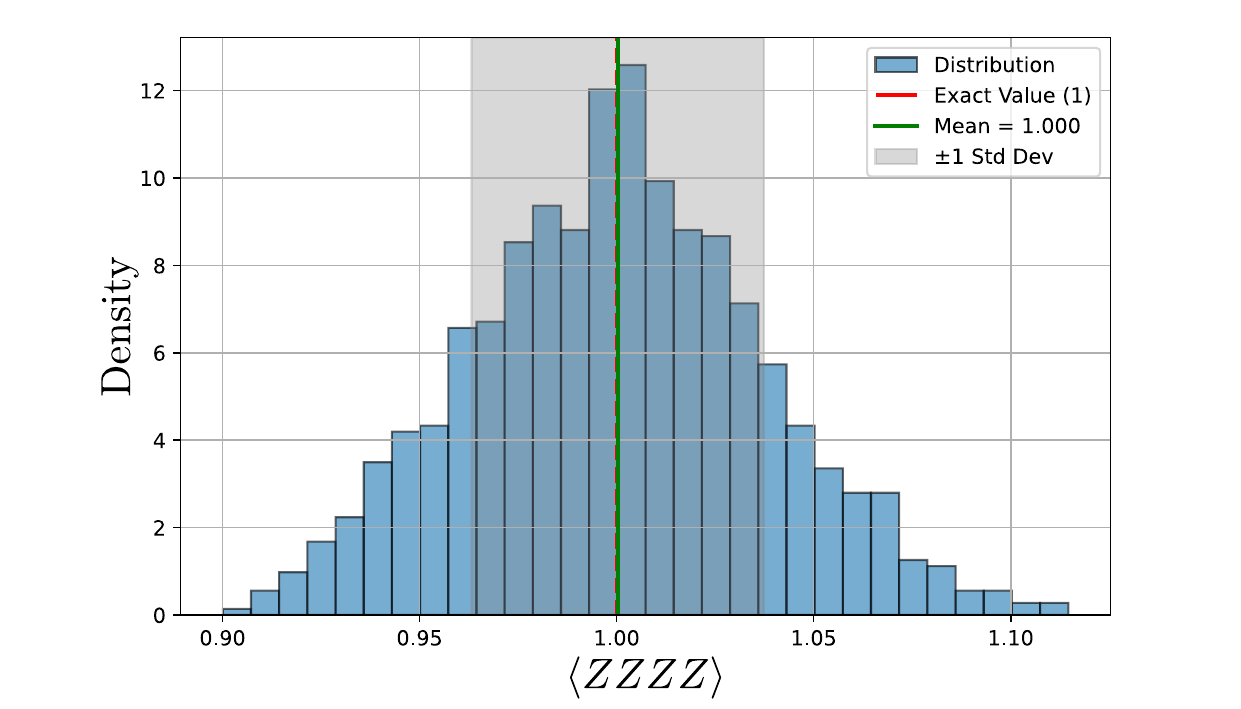}
    \caption{Validation of GHZ preparation using QDP: Distribution of the computed values of $\langle ZZZZ \rangle$, obtained after 1000 repetitions, showing the mean value and standard deviation. The mean of the results is approximately 1 (green line), closely matching the exact expected value 1 (red dashed line). The gray-shaded region represents the standard deviation of the data. This demonstrate the successful approximated preparation of the GHZ using QDP.}
    \label{fig:validation}
\end{figure}

Some of the coefficients $c_k$ are negative, necessitating the introduction of an overhead factor $\gamma = \sum_k |c_k|$ and a weight $  |c_k|/\gamma$ to accurately restore the original probability distribution. The overhead factor, $\gamma$, quantifies the deviation of the QPD from the original distribution or mean value. We focus on the mean value $\langle ZZZZ \rangle$, which is approximated by the estimator as follows, based on (\ref{eq:qdp}):
\begin{eqnarray*}
\gamma^2 \sum_{k,s} \frac{|c_k|}{\gamma}\frac{|c_s|}{\gamma} \sum_{\substack{(y_1, ..., y_4),\\(o_s, o_k)}}\text{sgn}(c_k)\text{sgn}(c_s) o_k o_s f(y_1, y_2, y_3, y_4) \\ 
\times P[y_1, o_k | k] \cdot P[y_2, o_s | k] \cdot P[y_3, y_4 | s], 
\end{eqnarray*}
where $P[y_1, o_k | k]$, $P[y_2, o_s | k]$, and $P[y_3, y_4 | s]$ represent the probability distributions conditioned on indices $k,s$, and $o_{k}$, $o_{s}$ are the measurements of ancillary qubits (see Fig. \ref{fig:ghz}b). For more details about the probability distributions calculation and wire cutting in general see~\cite{harada2024doubly}. The function is a map from 
$\{0,1\}^N$ to $\{-1,1\}$, transforming specific vectors such as $(y_1 = 0, y_2 = 1, y_3 =1, y_4 = 0)$ into $-1$ or $1$ (e.g., $f(y_1,y_2,y_3,y_4)=(2y_1-1)(2y_2-1)(2y_3-1)(2y_4-1)$). In a simulated experiment conducted with 1024 shots (number of quantum measurements) per execution, using 64 cores, we get an approximated value for $\langle ZZZZ \rangle \approx 1$, which aligns well with the theoretical expectation 1. To validate the correctness of our implementation, we analyze the distribution of computed values for $\langle ZZZZ \rangle$ obtained after 1000 repetitions (see Fig. \ref{fig:validation}). The mean of the results is approximately 1, confirming agreement with the exact expected value. The gray-shaded region represents the standard deviation of the data, which can be reduced by increasing the number of shots~\cite{harada2024doubly}.

\section{Conclusion}
\label{sec:outlook}
As heterogeneous systems continue to evolve and new computing technologies rapidly emerge, integrating quantum computing with classical task-based runtimes presents both opportunities and challenges. This paper outlines one potential approach for achieving such integration by leveraging the IRIS task-based runtime, XACC, qiskit and QIR-EE execution engines, enabling execution of hybrid classical-quantum programs. The work shown here demonstrates potential for Q-IRIS to support complex hybrid workflows. This was illustrated in Section~\ref{sec:ghz} with the parallel decomposition and execution of a four-qubit GHZ quantum state preparation task utilizing the integration described in Section~\ref{sec:qiris}. This highlights the \textit{feasibility} of integrating quantum computing with heterogeneous, many-task execution models and allows quantum simulations to coexist with distributed computing. 

Further developments will extend Q-IRIS to support workflows that leverage genetic algorithms for variational quantum solvers~\cite{alvarez2023gene}, parallel execution of quantum instances for gradient-free optimization methods~\cite{leyton2021robust,nguemto2022re}, and distributed quantum execution where multiple QPUs communicate via classical channels~\cite{carrera2024combining,harada2024doubly}. Additionally, the exploration of MLIR~\cite{MLIR2021,nvgpuMLIR,CatalystMLIR} to generate optimized codes that can be lowered to QIR for seamless interoperability between classical and quantum computing resources, is another natural direction of research for discovering the future capabilities of Q-IRIS.

We imagine that the scientific impact of Q-IRIS spans quantum machine learning, variational quantum algorithms, and high-performance quantum simulations. By presenting an efficient execution model, Q-IRIS has the potential to significantly contribute to bridging the gap between classical and quantum computing paradigms. Before that happens, key challenges will need to be addressed including: latency in classical-quantum communication, scheduling overhead in managing heterogeneous resources, and error mitigation strategies for improving execution reliability. These improvements will scale Q-IRIS capabilities for larger, more complex hybrid workflows.

\begin{acks}
    Notice: This manuscript has been authored by UT-Battelle LLC under contract DE-AC05-00OR22725 with the US  Department of Energy (DOE). The US government retains and the publisher, by accepting the article for publication, acknowledges that the US government retains a nonexclusive, paid-up, irrevocable, worldwide license to publish or reproduce the published form of this manuscript, or allow others to do so, for US government purposes. DOE will provide public access to these results of federally sponsored research in accordance with the DOE Public Access Plan (https://www.energy.gov/doe-public-access-plan).

    This work was supported by the U.S. Department of Energy, Office of Science under Contract No. DE-AC05-00OR22725, with funding from the Office of Advanced Scientific Computing Research's Accelerated Research in Quantum Computing Program MACH-Q project and resources of Oak Ridge National Laboratory's Experimental Computing Laboratory (ExCL).
\end{acks}

\bibliographystyle{ACM-Reference-Format}
\bibliography{bibliography}

\end{document}